\documentclass[12pt]{article}

\usepackage{amssymb}
\usepackage{latexsym}

\def\fnote#1#2{\begingroup\def\thefootnote{#1}\footnote{#2}\addtocounter{footnote}{-1}\endgroup}

\def\inbar{\vrule height1.5ex width.4pt depth0pt}
\def\IB{\relax{\rm I\kern-.18em B}}
\def\IC{\relax\,\hbox{$\inbar\kern-.3em{\rm C}$}}
\def\ID{\relax{\rm I\kern-.18em D}}
\def\IE{\relax{\rm I\kern-.18em E}}
\def\IF{\relax{\rm I\kern-.18em F}}
\def\IG{\relax\,\hbox{$\inbar\kern-.3em{\rm G}$}}
\def\IH{\relax{\rm I\kern-.18em H}}
\def\II{\relax{\rm I\kern-.18em I}}
\def\IK{\relax{\rm I\kern-.18em K}}
\def\IL{\relax{\rm I\kern-.18em L}}
\def\IM{\relax{\rm I\kern-.18em M}}
\def\IN{\relax{\rm I\kern-.18em N}}
\def\IO{\relax\,\hbox{$\inbar\kern-.3em{\rm O}$}}
\def\IP{\relax{\rm I\kern-.18em P}}
\def\IQ{\relax\,\hbox{$\inbar\kern-.3em{\rm Q}$}}
\def\IR{\relax{\rm I\kern-.18em R}}
\def\IT{\relax{\rm I\kern-.18em T}}
\def\ZZ{\relax{\sf Z\kern-.4em Z}}

\DeclareFontEncoding{OT2}{}{} 
  \newcommand{\textcyr}[1]{%
    {\fontencoding{OT2}\fontfamily{wncyr}\fontseries{m}\fontshape{n}%
     \selectfont #1}}
\newcommand{\Sha}{{\mbox{\textcyr{Sh}}}}

       \def\g{\gamma}  
\def\e{\epsilon} \def\G{\Gamma}     \def\l{\lambda}
\def\L{\Lambda}  \def\om{\omega}  \def\Om{\Omega} \def\si{\sigma}
\def\Si{\Sigma}       

\def\cC{{\cal C}} \def\cD{{\cal D}}  
 \def\cH{{\cal H}}  
  \def\cM{{\cal M}} \def\cN{{\cal N}}
\def\cO{{\cal O}}

\def\mathC{{\mathbb C}}  \def\mathF{{\mathbb F}}
\def\mathN{{\mathbb N}}  \def\mathP{{\mathbb P}} \def\mathQ{{\mathbb Q}}
\def\mathR{{\mathbb R}}  \def\mathZ{{\mathbb Z}}

\def\tL{{\tilde L}}

\def\hhat{{\hat h}}

\def\oK{{\overline K}}

\def\fnote#1#2{\begingroup\def\thefootnote{#1}\footnote{#2}\addtocounter
{footnote}{-1}\endgroup}
\def\beq{\begin{equation}}
\def\eeq{\end{equation}}
\def\bea{\begin{eqnarray}}
\def\eea{\end{eqnarray}}

\def\lleq#1{\label{#1}\eeq}
\let\nn=\nonumber
\def\tabroom{\hbox to0pt{\phantom{\Huge A}\hss}}
\def\notin{\ \hbox{{$\in$}\kern-.51em\hbox{/}}}

\def\lra{\longrightarrow}

\def\ra{{\rightarrow}}

  \def\E1Fq{E_1/\IF_q}

\def\rmD{{\rm D}}

 \def\rmdet{{\rm det}}   \def\rmdim{{\rm dim}}
          \def\rmgen{{\rm gen}}
\def\rmhom{{\rm hom}}

\def\rmker{{\rm ker}}
\def\rmmax{{\rm max}}   \def\rmmod{{\rm mod}}
     \def\rmord{{\rm ord}}
\def\rmrk{{\rm rk}}
 
 \def\rmtor{{\rm tor}}
\def\rmtr{{\rm tr}}

   \def\rmBSD{{\rm BSD}}
\def\rmCFT{{\rm CFT}}  \def\rmCH{{\rm CH}}    
       \def\rmGal{{\rm Gal}}

\def\rmI{{\rm I}}       \def\rmII{{\rm II}}

     \def\rmRe{{\rm Re}}
\def\rmSh{{\rm Sh}}       
\def\rmSL{{\rm SL}}

\def\notdiv{{\relax{~|\kern-.35em /~}}}

\def\boxit#1{
\vbox{\hrule height1pt\hbox{\vrule width1pt\kern0.3cm
\vbox{\kern0.3cm\hbox{$\displaystyle#1$}\kern0.3cm}\kern0.3cm\vrule
width1pt}\hrule height1pt}}

\begin{document}
 \parindent=0pt


\hfill CERN$-$PH$-$TH/2012$-$118


 \vskip 1truein

 \centerline{\Large {\bf $K-$Rational D$-$Brane Crystals}}

\vskip .3truein

 \centerline{\sc Rolf Schimmrigk\fnote{$\dagger$}{Email: netahu@yahoo.com; rschimmr@iusb.edu. 
    On leave of absence from  Indiana University South Bend, USA}}

\vskip .2truein

\centerline{Theory Division, CERN}
   \vskip .05truein
\centerline{CH$-$1211 Geneva 23, Switzerland}



\vskip 1truein

\baselineskip=17pt

\centerline{\bf ABSTRACT}

\vskip .1truein

\begin{quote}
 In this paper the problem of constructing spacetime from string
 theory is addressed in the context of D$-$brane physics. It is 
 suggested that the knowledge of discrete configurations
 of D$-$branes is sufficient to reconstruct the motivic building blocks of
  certain Calabi-Yau varieties. The collections of D$-$branes
 involved have algebraic base points, leading to the notion of
 $K-$arithmetic D$-$crystals for algebraic number fields $K$.
 This idea can be tested for D0$-$branes in the framework of toroidal
 compactifications via the conjectures of Birch 
and Swinnerton-Dyer. For the special class of D0$-$crystals of
 Heegner type these conjectures can be interpreted 
 as formulae that relate the canonical N\'eron-Tate height of the base points 
of the D$-$crystals to special values of the motivic L$-$function 
 at the central point.  In simple cases the knowledge of the D$-$crystals of Heegner 
type suffices to uniquely determine the geometry.
\end{quote}

\vskip .3truein


\renewcommand\thepage{}
\newpage


 \parindent=0pt
 \pagenumbering{arabic}

 \baselineskip=21pt

 \tableofcontents

 \vfill \eject

 \baselineskip=20.7pt
 \parskip=.2truein

\section{Introduction}

It has become apparent over the past few years that motivic
$L-$functions encode interesting physical information. In the
context of Calabi-Yau manifolds and their natural generalizations
there is a distinguished $L-$function, associated to the
$\Om-$motive of a variety $X$, that has been shown to be modular in
examples that cover all physically relevant dimensions. In these
instances the $L-$function leads to modular forms which can be
expressed in terms of modular forms of the corresponding conformal
field theory on the worldsheet, leading to relations of the type
 \beq
  f_{\Om}(X,q) ~=~ \prod_i \Theta^{k_i}_{\ell_i,m_i}(q^{a_i}) \otimes
   \chi_K.
 \lleq{bp-modularity}
 Here the Mellin transform of the cusp form $f_{\Om}(X,q) \in S_w(\G_0(N),\e_N)$ 
is the $L-$function of  the $\Om-$motive $M_{\Om}$ of $X$, the functions
 $\Theta^{k_i}_{\ell_i,m_i}(q)$, with $q=e^{2\pi i \tau}$, are
 Hecke indefinite modular forms, and $\chi_K$ is the quadratic character
 associated to a number field determined by the conformal field
 theory. These results show that the
arithmetic information of an exactly solvable variety carries the
essential information contained in the worldsheet model. Several examples 
of ths type have been constructed in different dimensions in 
 \cite{su02,rs06,rs08} for diagonal varieties, and ref. \cite{kls10}  for families 
 of manifolds. Viewing this relation as a map from the worldsheet conformal 
field theory $\rmCFT_\Si$ to the Calabi-Yau variety $X$ gives meaning to the notion 
of an emergent spacetime in string theory as a direct construction
 $\rmCFT_\Si \lra X$ via automorphic motives.

It is natural to ask whether motivic $L-$functions contain
information about the string vacuum other than the worldsheet
modular forms, in particular whether there are physical objects in spacetime 
that are described by the $L-$function. The purpose of this paper
is to address this question by showing that the $L-$function
contains information about particular types of D$-$branes on the
compactification variety, and that these  D$-$branes in turn contain enough
information to identify the basic building blocks of the compact
manifold. Roughly speaking, the $L-$function measures the detailed
structure of D$-$branes via the canonical height of the D$-$branes.  
This leads to a notion of D$-$brane
 multiplicities, and in the process leads to a direct 
 spacetime interpretation of the $L-$function.

The concrete definition of the 'arithmetic'
D$-$branes adopted here is motivated by the results of \cite{rs01},
where it was shown that certain algebraic number fields
have both a geometric and a conformal field theory interpretation.
This suggests that a natural probe for the structure of spacetime is provided
by D$-$branes that are defined over complex algebraic number fields $K$,
i.e. finite extensions of the rational numbers $\mathQ$. In this framework D$-$branes
 that wrap around cycles $C$ of the manifold are viewed as objects that are defined 
 over the extensions $K/\mathQ$. This construction should not necessarily be viewed as 
a replacement of the archimedean fields common in physics, but instead should be 
interpreted as a tool to probe the domain of  D$-$branes in various ways, 
depending on the nature of the field $K$. The structure of $C-$wrapped D$-$brane defined 
over $K$ will be denoted by $\cD_K(C)$ and called $K-$rational D$-$crystals. More generally
 D$-$crystals are associated to all arithmetic cycles, denoted by $\cD_K(\cC)$, where
 $\cC$ is the set of all cycles. 

The simplest framework in which the idea of relating D$-$brane probes to L-functions can be made 
precise and tested is given by D0$-$branes in toroidal compactifications. 
In this case the moduli space$\cM_{\rmD 0}(X)$ of D0$-$branes on a variety $X$ is the manifold
$X$ itself, $\cM_{\rmD 0}(X)=X$. Put differently, in the case of D0$-$branes there are no non-trivial 
cycles that can be wrapped and in the arithmetic setting 
the base points of the D0$-$branes can be located 
 at all $K-$rational points of the variety, leading to $\cD_K(X) = X(K)$. 
 In the general the connection between the $K-$arithmetic D$-$crystals and the $L-$function 
 is based on one of the central themes in arithmetic geometry, the conjecture of 
 Birch and Swinnerton-Dyer and its generalizations, the Bloch-Kato conjecture. 
  For D0$-$crystals on elliptic curves the key measure  with which they can be characterized
 in a numerically precise way is given by the N\'eron-Tate canonical height.
  The conjecture of Birch and Swinnerton-Dyer (BSD conjecture \cite{bsd63, bsd65}), relates this
 height, in combination with other factors, to the motivic $L-$function
evaluated at certain critical points. Very roughly, the conjectured 
relation 
  \beq
  L^{(r)}(E,1) ~=~ b_E \Om_E R_E
  \eeq
  links the Taylor coefficient $L^{(r)}(E,s)$ of the elliptic curve $E$
  at a special point $s=1$ to the regulator $R_E$ constructed from the
  height function and the period $\Om_E$. 
Here $b_E$ is a rational factor which will be made
  explicit later in this paper.
  The $L-$function in turn can be used to reconstruct
(motives of) the compactification manifold, and determine string
worldsheet modular forms. In this way it is possible to determine
the structure of spacetime via D0$-$brane physics.

For arbitrary complex number fields $K$ it is difficult to determine the structure 
 of the base points of the associated D$-$crystals because no constructive 
procedure is known at present.  This, in combination with the results of 
 \cite{rs01}, motivates the consideration of a subclass 
of $K-$algebraic D$-$crystals whose base points take values in number fields that 
are more general than the rational numbers, but special enough to be under 
control. This class of fields is given by the sequence of imaginary quadratic 
fields $K_D := \mathQ(\sqrt{-D})$ of discriminant $-D$.  For elliptic curves
 over $K_D$ a BSD type theorem has been proven by Gross-Zagier \cite{gz86} . Their
results can be interpreted as providing an explicit construction
of $K_D-$rational D0$-$branes via the notion of Heegner points $P_D$
associated to certain imaginary quadratic extensions $K_D$. For elliptic curves
of rank one over $K_D$ the N\'eron-Tate height $\hhat(P_K)$ of these objects
leads to the derivative of the $L$$-$function of the curve $E/K$
defined over $K$
 \beq
  L'(E/K,1) ~=~ c_{E,K}~\hhat(P_K),
 \eeq
 where $c_{E,K}$ is a number that depends on the geometry of $E$ and on $K$.
The relation between Heegner type D$-$branes and $L-$functions
established in this way is completely general, independent of any
knowledge of a worldsheet interpretation of the motivic
$L-$function. It therefore allows a motivic analysis via D$-$branes
for families of varieties, independently of the conformal field
theoretic structure. Using results from Faltings then makes it
possible to construct the curve $E$ up to isogeny from the inverse
Mellin transform of the $L-$function $L(E,s)$.


The outline of this paper is as follows. Sections 2 and 3 contain an
outline of the conjectures of Birch and Swinnerton-Dyer as well as a discussion 
of arithmetic D$-$branes. Section 4 explains the notion of
Heegner type D0$-$branes and Section 5 describes how they lead to motivic
$L-$functions that allow to reconstruct the compactification
geometry. Section 6 describes how Heegner type D$-$crystals determine 
the compactification variety and Section 7 contains two examples, one to illustrate the
constructions in the context of an exactly solvable elliptic
curve, the second to illustrate that the ideas here are more
general than the framework of exactly solvable compactifications.

\section{Rational D$-$crystals and $L-$functions}

The idea that some special sub-class of D$-$branes should encode
sufficient information to reconstruct the (motivic) structure of
spacetime leads to the question whether some particular class 
of D$-$branes can be related to motivic $L-$functions derived from a variety. In
this section this problem is addressed in the simplest possible
context, provided D0$-$branes on a toroidal manifold. This provides 
 a framework that is quite challenging mathematically, with many open 
conjectures, but simple enough to contain some proven results. It is also 
 a good starting point because it avoids the complications introduced in higher 
dimensions by the necessity to consider motives. In the elliptic case the situation 
is simpler because the question raised translates into the problem of 
whether there exists a relation between the Hasse-Weil
$L-$function $L(E,s)$ of an elliptic curve and particular types of
points on the variety which define the base points of the D$-$branes. 
When considering possible types of D0$-$brane base points as candidates it is 
natural to ask whether base points given by rational points, or some
subset thereof, provide suffiently sensitive probes of the spacetime manifold. 
This question motivates a closer investigation of the relation between rational 
points on elliptic curves and
$L-$functions.

\subsection{Rational D$-$crystals on tori}

The problem of understanding the rational points on abelian
varieties in general, and of elliptic curves in particular, is a
quite difficult one that has a long history. In the context of
elliptic curves the first structural result is the theorem of
Mordell \cite{m22}, which says that the group of rational points
$E(\mathQ)$ is finitely generated as
  \beq
  E(\mathQ) ~\cong ~ \mathZ^r \times E(\mathQ)_{\rmtor},
 \eeq
 where $r$ is called the rank of the group (or elliptic curve) and
 $E(\mathQ)_{\rmtor}$ denotes the points of finite order. The torsion points
 can be determined algorithmically and the possible groups are known by
 a theorem of Mazur \cite{m77}, which proves that a previously established
 list of groups is complete \cite{o75}. No such algorithm is known for the free part of the
 Mordell-Weil group. It is in particular not known whether the
 rank is bounded.

A connection between the structure of the Mordell-Weil group and the
$L-$function has not been proven in full generality at this point, but 
partial results are known, confirming a rather detailed picture of such a relation.
  The precise form of this link involves several other quantities that characterize 
the arithmetic structure of the curve and is formulated in two deep conjectures by 
 Birch and Swinnerton-Dyler that 
 basically relate the Taylor series structure of the $L-$series at the critical point 
 to  numerical information associated to the Mordell-Weil group, mediated by 
 further arithmetic information \cite{bsd63,bsd65}.

 Geometric $L-$functions can be viewed as tools that characterize the fine
 structure of a manifold. In the case of an elliptic curve $E$
 of discriminant $\Delta$ the $L-$function can be defined via the Euler
 product
 \beq
 L(E,s) = \prod_{p|\Delta} (1-a_pp^{-s})^{-1} \prod_{p\notdiv
 \Delta}  (1-a_pp^{-s} + p^{1-2s})^{-1},
 \lleq{l-function}
 where the coefficients $a_p$ are given by $a_p = p+1 -N_p$, where
 $N_p = \#(E(\mathF_p)$ is the number of points of $E$ over the
 finite field $\mathF_p$. A bound determined by Hasse for the
 coefficients shows that this function converges in the half-plane
 $\rmRe(s) >3/2$. In order to formulate the conjectures of Birch
 and Swinnerton-Dyer it is necessary to evaluate the $L-$function at
 $s=1$. The resulting problem of analytic continuation was solved by Wiles et al.
  \cite{w95, bcdt01} in the context of the proof
 of the Shimura-Taniyama-Weil conjecture. According to the resulting elliptic modularity
 theorem there exists a modular form $f\in S_2(\G_0(N))$ such that the 
$L-$function is essentially given by the Mellin transform, more precisely
 \beq
     \int_0^{\infty} f(iy) y^{s-1} dy = (2\pi)^{-s} \G(s) L(f,s).
 \eeq
 It was shown by Carayol that $f$ is a newform and an
 eigenfunction of the Fricke involution 
  \beq
  f(-1/N\tau) = \l N\tau^2f(\tau),
  \eeq
 where $\l = \pm 1$.
  This transformation behavior motivates the introduction of a renormalized 
 completed $L-$function as 
  \beq
   L^*(f,s) ~:=~ N^{s/2} \frac{\G(s)}{(2\pi)^s} ~L(f,s),
  \lleq{completed-L-function}
   which satisfies the functional equation
  \beq
     L^*(f,2-s) = \e L^*(f,s),
 \lleq{functional-equation}
 where $\e = -\l \in \{\pm 1\}$. This functional equation shows that the
 point $s=1$ is of particular significance since $L^*(f,1)=\e L^*(f,1)$.
 Thus if $\e=-1$ the $L-$function necessarily vanishes.

\subsection{The Birch$-$Swinnerton-Dyer conjectures}

The idea described above of probing the spacetime geometry by D$-$branes and
relating their structure to the worldsheet physics via their associated $L-$functions
 translates into the question whether the base points of
the D$-$crystals can be related to the motivic $L-$function. The
rational positioning of the D$-$crystal leads to the Mordell-Weil
group, hence the $\mathQ-$rational base points lead to the
conjectures of Birch and Swinnerton-Dyer.

 The difficulties that arise in the problem of understanding the 
  Mordell-Weil group $E(\mathQ)$ led Birch and Swinnerton-Dyer in the 1960s 
  to  perform extensive computer computations that suggested relations between 
  the $L-$function and certain other characteristics that encode the arithmetic 
 structure of elliptic curves.
 The first part of these  conjectures \cite{bsd63,bsd65} relates the rank $\rmrk~E(\mathQ)$ 
 of the Mordell-Weil group $E(\mathQ)$ to the vanishing order $\rmord_{s=1}L(E,s)$ 
   of the $L-$series $L(E,s)$ at the central critical
 point $s=1$. This is one of the Millenium Prize Problem established by the Clay Mathematics 
 Institute \cite{w06}.

\underline{BSD rank conjecture.}
 \beq
   \rmBSD_{\mathQ}\rmI:~~~~ \rmord_{s=1}L(E,s) = \rmrk~ E(\mathQ).
 \eeq

An intuitive observation concerning the structure of the $L-$function motivates this conjecture. 
 If one were to evaluate the $L-$function (\ref{l-function}) at $s=1$, the second factor, denoted by 
 $\tL(E,s)$,  takes the form
  \beq
  \tL(E,1) ~=~ \prod_{p\notdiv N} \frac{p}{p-a_p+1}.
  \eeq
 With $a_p= p+1-N_p(E)$, where $N_p(E):=\#E/\mathF_p$ is the cardinality of the elliptic curve over 
the finite field $\mathF_p$, this leads to 
 \beq
  \tL(E,1) ~=~ \prod_{p\notdiv N} \frac{p}{N_p(E)}.
  \lleq{bsd-intuitive}
 The expectation now is that if the Mordell-Weil group $E(\mathQ)$ is large then there are many points that 
 can be obtained via reduction by $p$ from the rational points. If many of the $N_p(E)$ are very large it is 
reasonable to expect $L(E,1)$ given by (\ref{bsd-intuitive}) to vanish.

 For elliptic curves with vanishing order $\leq 1$ the rank conjecture is known to hold.
 The arithmetic structure of elliptic curves however is much richer than just its rank 
 and it is  natural to ask whether the coefficients in the Taylor expansion of the
 $L-$function admit a geometric interpretation as well. The vision of Birch and Swinnerton-Dyer,
 guided by their experimental results, led to the conclusion that the $L-$function
 does in fact contain highly nontrivial information about the geometry. More
 precisely, the rank conjecture suggests to consider the Taylor
 expansion of the $L-$function around the critical point. If the
 rank of the Mordell-Weil group is abbreviated by $r=\rmrk~E(\mathQ)$
 this expansion takes the form
  \beq
   L(E,s) = d_r (s-1)^r + \cdots
  \eeq
  for some value $d_r$. The question then becomes whether one find a formula for $d_r$ 
 in terms of the geometry of $E$. This leads to the second part of the BSD conjectures, which can be
  formulated as follows.

 \underline{BSD Taylor conjecture.}
  \beq
 \rmBSD_{\mathQ}\rmII:~~~~
  \lim_{s\ra 1} \frac{L(E,s)}{(s-1)^r}
   ~=~ c_T \cdot \Om~
   \frac{|\Sha|}{|E(\mathQ)_{\rmtor}|^2}~R,
  \eeq
  where $c_T=\prod_p c_p$ is the product of the Tamagawa numbers, which
  are determined by the behavior of the curve at the bad primes as
  $c_p = [E(\mathQ_p):E_0(\mathQ_p)]$, with $E_0(\mathQ_p)$ the
  set of points that reduce to smooth points over $\mathF_p$. If $E$ has good reduction at $p$ then 
 $c_p=1$, hence $c_T$ is determined by a finite number of primes 
  \beq
  c^T ~=~ \prod_{p|\Delta(E)} c_p,
  \eeq
   where $\Delta(E)$ is the discriminant of $E$.
  The  period $\Om$ of an elliptic curve given via its
   generalized Weierstrass equation
   \beq
    y^2 + a_1xy + a_3y = x^3 + a_2x^2+a_4x + a_6
   \eeq
   is defined via the N\'eron differential
    \beq
     \om = \frac{dx}{2y+a_1x+a_3}
   \eeq
   as
   \beq
   \Om = \int_{E(\mathR)} \om.
  \eeq
   Very little is known about
   the Tate-Shafarevich group  $\Sha$. It
    is conjectured to be finite, but can be arbitrarily large. It
   can be shown \cite{c64} that for some fixed $c>0$ and
   infinitely many $N\in \mathN$ there exists an elliptic curve
   such that $\Sha(E) >> N^{c/\log \log N}$. This group is a
   torsion group which measures the failure of the Hasse
   principle, i.e. it detects whether a variety admits local
   points but no global points. It can be defined in terms of
   cohomology groups associated to the absolute Galois group
   $\rmGal(\oK/K)$ of any number field $K$. With the short-hand notation 
  $H^1(K,E) = H^1(\rmGal(\oK/K),E)$ the Tate-Shafarevich group is given 
   by
   \beq
   \Sha(E/\mathQ) = \rmker\left(H^1(\mathQ,E)  ~\lra ~
          \prod_v H^1(\mathQ_v,E)\right),
   \eeq
   where $\mathQ_v$ is the completion of $\mathQ$ at the place
   $v$. It turns out that this group is the most difficult part  to compute
   and much effort has gone into attempts to develop algorithms that
   allow to determine it.  
  For elliptic curves of rank $\leq 1$ this group is known to be finite, but 
 nothing is known for higher rank curves.  
  While mathematically difficult, the Shafaravich-Tate
  group is in many examples either trivial or small, hence it is not a particularly sensitive 
  characteristic of elliptic curves and will not play an essential role in this paper. 

  It will become clear further below that one of the most relevant quantities
   in the present discussion is the regulator $R_E$ of the
  elliptic curve. This quantity is defined as the determinant of a height
  pairing defined in terms of the N\'eron-Tate
  height of the generators $P_i$ defining a basis
  of the Mordell-Weil group of $E$.  More precisely,
  the N\'eron-Tate canonical height of a rational point is defined
  in terms of the logarithmic height $h(P)$ of a point 
   $P=(x,y) \in E$. Writing $x=r/s$ where $r,s \in \mathZ$ have no common 
  factor, the latter is defined as
   \beq
    h(P) = \log~\rmmax\{|r|,|s|\},
   \eeq
   leading to the definition of the N\'eron-Tate height as
   \beq
    \hhat(P) = \lim_{n\ra \infty} \frac{h(2^nP)}{4^n}.
   \eeq
   This function satisfies the homogeneity condition $\hhat(nP) = n^2\hhat(P)$ and 
it vanishes if and only if the point $P$ is a torsion point. 
   The N\'eron-Tate height in turn can be used to define the height pairing
  \beq
    \langle P,Q\rangle = \frac{1}{2}\left(\hhat(P+Q) -
    \hhat(P)-\hhat(Q)\right)
   \eeq
   which defines a  nondegenerate real quadratic form on the free part of the Mordell-Weil group 
   $E(\mathQ)/E(\mathQ)_{\rmtor}$ because of the triviality of the N\'eron-Tate height on the torsion part.
  The regulator $R_E$, finally, is defined as
  \beq
   R_E = \rmdet (\langle P_i,P_j\rangle)_{i,j=1,...,\rmrk(E))}.
  \eeq

 An intuitive physical interpretation of the logarithmic height is in
 terms of the complexity of D0$-$brane in the sense of
 measuring the amount of energy, or time, needed to communicate
 the structure of the algebraic points at which the D0$-$branes are
 located. The conjecture of Birch and Swinnerton-Dyer thus relates
 the complexity of D0$-$brane crystals to the critical values of the motivic
 $L-$function.

The BSD conjectures have been proven for certain types of elliptic
curves, which will be made more explicit further below. They provide the
first realization of the sought-after link between a certain class
of base points of D0$-$branes on the simplest possible
compactification. With this framework in place one can ask whether
rational D0$-$branes lead to interesting information about the
varieties on which they live.

\subsection{String modular examples of rank 0 and torsion D$-$crystals}

With the conjectures of Birch and Swinnerton-Dyer in place it is possible to check 
what the relations BSD$_\mathQ$I and BSD$_\mathQ$II imply in the context of a special class of string 
compactifications given by exactly solvable elliptic curves of Gepner type. For these 
models the underlying conformal field theory is a tensor product of $\cN=2$ 
superconformal minimal models whose partition function is described by 
characters that are constructed from Kac-Petersson string functions of the 
underlying $A_1^{(1)}$ affine Lie algebra. These string functions in turn 
are determined by Hecke indefinite modular forms and it was shown 
in ref. \cite{rs05} that the Mellin transform of twisted products of these 
Hecke indefinite modular forms are identical to the $L-$function of the
corresponding elliptic curves via equations of type
(\ref{bp-modularity}). These results show that the arithmetic
structure of the compactification manifold encodes the physics on
the worldsheet without the need to probe the geometry over
continuous fields, such as $\mathR$ or $\mathC$. The link
 between the worldsheet physics and the spacetime geometry is provided
 by the $L-$function and its associated modular form. This provides
 a first physical interpretation of the $L-$function.

The detailed structure of the modular forms $f(E,q)$ such that
$L(f,s) = L(E,s)$ for these three diagonal elliptic curves involve
Hecke indefinite modular forms $\Theta^k_{\ell,m}$, where $k$
indicates the level of the conformal field theory and $\ell,m$ are
quantum numbers within the CFT. The results are summarized in
Table 1.
 \begin{center}
\begin{tabular}{l| c  c}

Curve $E_d$              &Worldsheet representation   &Space  \tabroom \\

\hline

 $E_3 \subset \mathP_2$ &$f(E_3,q) ~=~ \Theta^1_{1,1}(q^3)\Theta^1_{1,1}(q^9)$
       & $S_2(\G_0(27))$  \tabroom \\

 $E_4 \subset \mathP_{(1,1,2)}$ &$f(E_4,q) ~=~\Theta^2_{1,1}(q^4)^2\otimes \chi_2$
       &$S_2(\G_0(64))$  \tabroom \\

 $E_6 \subset \mathP_{(1,2,3)}$ &$f(E_6,q) ~=~ \Theta^1_{1,1}(q^6)^2\otimes \chi_3$
      &$S_2(\G_0(144))$  \tabroom \\

\hline
\end{tabular}
\end{center}

\centerline{{\bf Table 1.}~{\it Modular $L-$functions of diagonal
elliptic curves \cite{rs05}.}}

The exactly solvable models listed in Table 1 have rank 0, hence the base points of the 
$\mathQ$-rational D$-$crystals are of torsion type. Since the height function is trivial on the 
torsion factor of the Mordell-Weil group the regulator is trivial and is set to $R=1$. The conjecture 
of Birch and Swinnerton-Dyer therefore specializes for these models to
 \beq
  L(E,1) = c_T\cdot \Om ~\frac{|\Sha|}{|E(\mathQ)_{\rmtor}|^2}.
 \lleq{bsd2-rk0}
 As mentioned above, the Tamagawa number $c_T$ receives
 contributions only from the bad primes $p|N$, where $N$ is the
 conductor of the curve. For the examples here these conductors
 are given by $N=27,64,144$, respectively, hence there are at most
two bad primes.  The Brieskorn-Pham curves listed in Table 1 admit complex multiplication 
 over some imaginary quadratic field $K$ for which $L(E/K,1)\neq 0$. For CM curves Rubin  
 has shown that the Shafarevich-Tate group $\Sha_K(E)$ is finite and that the 
 BSD conjecture holds \cite{r87}. It is therefore possible to compute the order of 
the Shafarevich-Tate group by using the BSD II relation (\ref{bsd2-rk0}).  
 For the exactly solvable elliptic curves of Brieskorn-Pham type
 the numerical results collected in Table 2 show that the formula of Birch and
 Swinnerton-Dyer leads to trivial Shafarevich-Tate group
 $|\Sha|=1$ for all three curves.
 \begin{center}
 \begin{tabular}{c| c c c c c c  c}

 Elliptic curve  &$L(E,1)$   &$\Om$    &$|E(\mathQ)_{\rmtor}|$
                 &$\prod_p c_p$       &$r$  &$|\Sha|$  \tabroom \\
 \hline

 $E_{27}$    &0.588879...     &1.766638...  &3   &3    &0  &1 \tabroom \\

 $E_{64}$    &0.927037...     &3.708149...  &2   &1   &0  &1 \tabroom \\

 $E_{144}$   &1.214325...     &2.428650... &2   &$1\cdot 2$  &0 &1 
 \tabroom \\

 \hline
 \end{tabular}
 \end{center}

 \centerline{{\bf Table 2.}~{\it Birch-Swinnerton-Dyer data
   for the elliptic curves $E_N$ analyzed in \cite{rs05}.}
   }

It becomes clear from the results collected in Table 2 that the Shafarevich-Tate group, while the
 most difficult to compute, is not a very sensitive characteristic, a result 
that is confirmed by the data base developed by Cremona. Essentially then,
 the BSD conjectures  show that in the case of exactly solvable
  curves the $L-$functions of the modular forms
 on the worldsheet evaluated at the central value contain 
  global information about the compactification geometry,
 such as the periods, and the torsion points.
 More precisely,  for the examples summarized in Tables 1 and 2 the data shows in
particular that the quotient $L(E,1)/\Om_E$ of the $L-$function evaluated at the
critical point by the period determines the size of the group of the base points of the 
$\mathQ-$rational D$-$crystal of torsion type on the curve. Since the period itself is
determined by the modular form it follows that the order of the
group of torsion type D0$-$branes is determined completely by the
modular form derived from the worldsheet  conformal field theory. 
Thus the worldsheet modular form determines the size of the rational D$-$crystal 
  $\cD_\mathQ(E)$.

More generally, for elliptic curves with complex multiplication it was shown first 
by Coates and Wiles \cite{cw77} that $L(E,1)\neq 0$ implies that the rank
of the elliptic curve is zero.  Work by Gross-Zagier \cite{gz86}, Kolyvagin \cite{k88a,k88b,k90},
Bump-Friedberg-Hoffstein \cite{bfh89, bfh90} and Murty-Murty
\cite{mm91} generalizes the Coates-Wiles result to all modular
curves and furthermore shows that if $L(E,1)=0$ and $L'(E,1)\neq 0$  
 then the rank of $E$ is equal to 1.
 Combining this with the elliptic modularity theorem of
Wiles et. al. leads to the conclusion that for any curve $E$ with
$L(E,1)\neq 0$ the Mordell-Weil group $E(\mathQ)$ is finite and that for all elliptic 
curves with $L(E,1)=0$ and $L'(E,1)\neq 0$ the rank of the Mordell-Weil group is one.
 Thus the conjecture BSD$_\mathQ$I is proven for curves whose rank is at most one.

The analysis of exactly solvable compactification shows 
that while $\mathQ$-rational D$-$branes do provide the sought after 
   link between the worldsheet physics and spacetime D$-$branes, 
 this type of probe is not particularly sensitive.   
 As noted earlier,  the possible torsion subgroups  of 
the Mordell-Weil group have been classified in work by Ogg and Mazur, with the result that 
the number of different groups is very limited, in particular the order of these groups is bounded by 
  $|E(\mathQ)_\rmtor|\leq 12$.  
 As a result the torsion part of the Mordell-Weil group $E(\mathQ)_\rmtor$ is an invariant that is
 enormously degenerate on the space of all elliptic curve, hence pure torsion D$-$crystals are not 
particularity discriminating probes. This motivates the consideration of more general 
D$-$crystals with base points 
that are not restricted to have $\mathQ-$rational base point coordinates, 
 but instead live in more general algebraic number fields $K$ that are extensions of $\mathQ$.

\section{$K-$rational D$-$crystals}

The results described above concerning torsion type D$-$crystals and
their relation to the worldsheet physics raise the question
whether extensions to more general
algebraic $K-$rational D$-$branes might be useful as more sensitive
probes of the geometry of the compactification manifold than $\mathQ$-rational D$-$crystals.
 The fact that the diagonal curves considered in the previous section
 have complex multiplication
 suggests that a natural class of fields to consider are imaginary quadratic fields.
 The purpose of this section is to describe the more general notion
  of $K-$rational D$-$crystals, which in the next Section are shown to
  provide a physical spacetime interpretation of the $L-$function. Later
  in this paper it will become clear that in the case
of elliptic curves (and their derived toroidal compactifications)
 these $K-$rational D$-$crystals contain sufficient information to reconstruct the
compactification manifold itself. In the case of exactly solvable
varieties they also are linked to the theory on the worldsheet via
certain modular forms of the conformal field theory.  Basic
background for elliptic curves can be found in the books by
Silverman \cite{silverman1, silverman2}.

The shift from $\mathQ-$rational to $K-$rational D$-$crystals
necessitates the generalization of Mordell-Weil group to the group
of $K-$rational base points $E(K)$. This in turn implies that a
generalization of the BSD conjectures is necessary. Such a
formulation has been provided by Tate \cite{t66}. The focus of the
first BSD conjecture now is the relation between the rank of
$E(K)$ and the $L-$function of the curve $E/K$ considered over
$K$.
 The generalized rank conjecture takes the form
 \beq
 {\rm BSD}_K\rmI:
  ~~~\rmord_{s=1} L(E/K,s) ~=~ \rmrk_K E(K),
 \eeq
 and the generalized Taylor conjecture becomes
 \beq
 {\rm BSD}_K\rmII:~~~~~
               \frac{1}{r!} L^{(r)}(E/K,1)
                 ~=~ c \cdot \Om_{E/K} ~
                   \frac{|\Sha(E/K)|}{|E(K)_{\rmtor}|^2}~R_{E/K},
 \eeq
 where $c=\left(\prod_p c_p\right)$ is again the Tamagawa number.

 The formulation of the general BSD conjectures for extensions
 $K/\mathQ$ of the rational numbers shows that it is possible in principle
 to address the question whether $K-$rational D$-$crystals for such number
 fields can be used to determine the underlying modular forms on the worldsheet.
  While it is difficult to make general statements about the structure of the
 Mordell-Weil group $E(K)$, it will become clear further
 below that it sufficient for the framework developed in this paper to
 consider extensions of $\mathQ-$rational D$-$crystals that are defined by
 imaginary quadratic fields $K_D= \mathQ(\sqrt{-D})$ of
 discriminant $-D$. For such fields a systematic construction of generators
 of $E(K_D)$ exists, based on the work of Birch and Gross-Zagier. This
 will lead to an infinite number of D$-$crystals defined over $K_D$ for running
 $D$, as described in the next section.

\section{D$-$crystals of Heegner type}

The discussion of  the $\mathQ-$rational D$-$crystals above shows that while these 
objects do contain information about the $L-$functions they are not sensitive enough 
to provide probes that are sensitive enough to contain all the information about the 
compactification geometry, i.e. they do not allow to identity isogeny classes of elliptic curves.
 It was also noted that general $K$-rational Mordell-Weil group $E(K)$ are very difficult to 
understand and to construct, hence are problematic as practical tools.  This motivates 
the search for types of D$-$crystals whose base points are more general than rational points, 
but are special enough to be under better control and which are rich enough to identify 
isogeny classes of elliptic curves. Such fields are in fact provided by the simplest type 
of generalization of rational numbers given imaginary quadratic fields
$K=\mathQ(\sqrt{-D})$ of discriminant $D$. This leads to the
question whether there exist non-torsion rational points when the
curve is considered over such fields. The problem of constructing
$K-$rational points is nontrivial and has first been considered by
Heegner in his solution of Gauss' class number problem \cite{h52}.
The question precisely when his construction leads to points of
infinite order has been discussed more systematically first by
Birch whose construction will be described in this Section.

The framework formulated in \cite{b75, bs84} is indirect,
starting with points on the modular curve $X_0(N) =
\overline{\cH/\G_0(N)}$, followed by a projection via the modular
parametrization $X_0(N) \lra E_N$ to an elliptic curve of
conductor $N$ associated to a modular form $f\in S_2(\G_0(N))$.
A second issue that arises is that the field of definition of the points obtained 
 on the elliptic curve via the modular parametrization map are defined not over 
the imaginary quadratic field $K_D$ but over its ring class field $H_D$. This necessitates 
 a trace map that projects the coordinates from $H_D$ to $K_D$.

\subsection{Modular parametrization of elliptic curves and CM points}

It was shown by Wiles et.al.  \cite{w95, bcdt01} that for every
elliptic curve $E_N$ with conductor $N$ there exists a modular
curve $X_0(N) = \cH/\G_0(N) \cup \{\rm cusps\}$ and a modular
parametrization map
 \beq
  \Phi_f:~~X_0(N) ~~\lra ~~ E_N(\mathC) ~=~ \mathC/\L_f
 \eeq
that is determined by a modular cusp form $f(q) \in S_2(\G_0(N))$ of
 weight two and level $N$ with respect to Hecke's congruence
 subgroup defined by
 \beq
 \G_0(N) ~=~ \left\{\g =\left(\matrix{a &b \cr c &d\cr}\right) \in
 \rmSL(2,\mathZ) ~{\Big |}~ c\equiv 0(\rmmod~N)\right\}.
 \eeq

 The modular curves $X_0(N)$ are in general of higher genus
  $$
  g(X_0(N)) ~=~ \rmdim~S_2(\G_0(N))
  $$
  and the elliptic curve can be obtained via Shimura's quotient construction by choosing 
   an element $f(q)$ in the space of cusp forms.
  Expanding $f(q)$ as a Fourier series $f(q) = \sum_n
 a_n q^n$ with $q=e^{2\pi i \tau}$ allows to write the
 series expansion of the map $\Phi_f$ as
 \beq
  \Phi_f(\tau) ~:=~ - \int_\tau^{i\infty} \om_f ~=~ \sum_n \frac{a_n}{n} e^{2\pi i n z},
 \eeq
 where the differential form is defined $\om_f := 2\pi i f(z)dz$.
 The lattice $\L_f$ of the image is given by
 \beq
  \L_f ~:=~ \left\{ \int_p^{\g p} \om_f ~{\Big |}~ \g \in
  \G_0(N)\right\}.
 \eeq

 The map from a lattice realization $E_\L = \mathC/\L$
of an elliptic curve to its algebraic form $E^w$ is obtained via
the Weierstrass function
  \beq
  w_{\L}(z) = \frac{1}{z^2} + \sum_{\l \in \L\backslash \{0\}}
   \left(\frac{1}{(z-\l)^2} - \frac{1}{\l^2}\right)
   \eeq
   as
   \bea
    \mathC/\L ~ &\lra & E^w \nn \\
          z       &\mapsto & (w_\L(z),w'_\L(z)).
  \eea
 The functions $(x,y) = (w_{\L}(z),w'_{\L}(z))$ satisfy
 the algebraic relation
 \beq
 y^2 = 4x^3 - g_2(\L)x - g_3(\L)
 \eeq
 where the coefficients $g_2,g_3$ are given in terms of the
 defining lattice $\L$ as
 \bea
  g_2 &=& 60 \sum_{\l\in \L \backslash \{0\}} \frac{1}{\l^4} \nn \\
  g_3 &=& 140 \sum_{\l \in \L \backslash \{0\}} \frac{1}{\l^6}.
 \eea

 The modular form $f(q) \in S_2(\G_0(N))$ associated to the curve $E$
 has a geometric interpretation as a power series whose Fourier coefficients
 $a_p$ for primes $p$ essentially measure the number of points $E(\mathF_p)$ of the
  elliptic curve over the finite fields $\mathF_p$. This leads to the Hasse-Weil L-series
  \beq
  L(E,s) ~=~ \sum_n \frac{a_n}{n^{s}}
  \eeq
  which in turn leads to $f(E,q)$ via inverse Mellin transform.
 For any specific curve $E$ such a modular parametrization can be
 found by explicit computation and the general machinery of Wiles
 et. al. is not necessary.

\subsection{The construction of D$-$crystals of Heegner-Birch type}

In this section a class of $K-$arithmetic D$-$crystals is described
whose existence can be proven. These D$-$crystals are defined such
that their base points are given by the special class of Heegner
points associated to imaginary quadratic fields $K_D =
\mathQ(\sqrt{-D})$ with discriminant $-D$. For such fields
nontriviality criteria for $E(K_D)$ are known. Furthermore, Gross
and Zagier have proven a precise form of the conjecture of
Birch-Stephens \cite{bs84} concerning a relation between the
height of Heegner-Birch points and the derivative of the
$L-$function at the central point in the rank 1 case. This result
shows that in this case the derivative of $L(E/K_D,s)$ at $s=1$
controls the existence of nontrivial $K-$rational points and
thereby provides the sought-after relation between D$-$crystals and
$L-$functions.

 The first step in this construction is to consider special points in the upper half plane.
 A point $z\in \cH$ is called a complex multiplication point (CM
 point) if it is the root of a quadratic equation
 \beq
  Az^2 + Bz + C =0,
 \eeq
  with $A,B,C \in \mathZ$ and $-D = B^2 -4AC<0$. Such points are special because 
the $j-$function evaluates to algebraic instead of transcendental values on CM points.

The next step of the construction of Heegner points given by Birch
\cite{b75} selects points $z\in \cH$ of the upper half plane such that $z$ and $Nz$ satisfy
 quadratic equations with the same discriminant
 \beq
  D(z) = D(Nz).
 \lleq{discriminant-constraint}
 The motivation for this condition arises from the combination of two facts. The first is that 
in general the $j-$function applied to arbitrary $z\in \cH$ and 
$Nz$ defines values $(x,y) = (j(z),j(Nz))$ that are not 
independent but satisfy an equation $F_N(x,y)=0$, where the structure of the defining 
polynomial depends on the level $N$. $F_N$ defines an algebraic curve $Z_0(N)=\{F_N(x,y)=0\}$ 
which is singular in general and whose resolution determines the modular curve $X_0(N)$.
 The second fact is that the nature of the $j-$function values $j(z)$ depends on the
 number theoretic character of $z$. As noted above, if $z$ is a CM point in some imaginary quadratic field $K_D$ of 
conductor $-D$ then $j(z)$ is algebraic, more precisely it 
takes values in the ring class field $R_D$ of $K_D$. The key now is that 
$j(Nz)$ takes values in a different field, hence the point $(j(z),j(Nz))$ is an element of $X_0(N)(\mathC)$ 
but is not defined over the ring class field. The constraint (\ref{discriminant-constraint}) ensures that 
 $(j(z),j(Nz)) \in X_0(N)(R_D)$. This will be of importance below.
  
 The existence of points that satisfy the discriminant constraint (\ref{discriminant-constraint}) 
depends on the structure of $D$ relative to the level $N$. Solutions exist when the 
 discriminant can be written as a square mod $4N$
 \beq
  r^2 \equiv -D (\rmmod~4N).
 \eeq
 This congruence condition can be formulated in a more conceptual way by noting that for such imaginary 
quadratic extensions all prime divisors $p$ of the conductor $N$ are split or ramified, i.e.
 \beq
  \chi_D(p) ~=~ \left(\frac{-D}{p}\right)=1, ~~~{\rm for~primes}~p|N,
 \lleq{character-constraint}
 where $\chi_D$ denotes the Legendre symbol.
 This constraint is called the Heegner constraint, and the discriminants
 which satisfy this condition are called Heegner discriminants.
 
Given a solution $r$ one can define points in the upper
 half-plane $\cH$ as
  \beq
  z_{D,r} = \frac{-B+\sqrt{-D}}{2A} ~\in ~\cH,
 \eeq
 where $N|A$ and $B\equiv r(\rmmod~2N)$. If the constraint (\ref{character-constraint}) is not 
satisfied then there are no such points, and if it is then the set $\{z_{D,r}\}$ of such points 
 is invariant under  $\G_0(N)$, leading  to $h(D)$ orbits $[z_{D,r}] \in X_0(N)$
 in the modular curve $X_0(N)$. Here $h(D)$ denotes the class number of the field $K_D$.
 In order to lighten the notation the orbits $[z_{D,r}]$ will again be denoted by $z_{D,r}$.

 The Heegner points on the curve $X_0(N)$ can now be
 mapped to an elliptic curve $E_N$ of conductor $N$ via the modular parametrization $\Phi_f$ map 
 associated to a modular form $f\in S_2(\G_0(N))$. This leads to points on the torus
  $E_N = \mathC/\L_f$ defined by the period lattice $\L_f$ of the modular form $f$. The image
 points $\Phi_f(z_{D,r})$ do not live in the original field.
 It follows from the theory of complex multiplication that
 instead of taking values in the imaginary quadratic field $K_D$ they
  are defined over ring class fields $R_D$ associated to $K_D$.
 If $K_D$ satisfies the splitting condition of Birch (\ref{discriminant-constraint}) this
 extension is the Hilbert class field $H_D$, i.e. the maximal
 abelian unramified extension of $K_D$,
 and they are permuted amongst themselves by the Galois group
  $\rmGal(H_D/K_D)$, whose order is the same as the class number $h(D)$.
 Taking the trace of an image point
  over this Galois group
 \beq
  \rmtr~ \Phi(z_{D,r}) ~=~ \sum_{\si \in \rmGal(H_D/K_D)}
  \si(\Phi(z_{D,r})
 \eeq
 therefore leads to points on the elliptic curve that are defined over
 $K_D$. The trace is divisible by $u_D =\frac{1}{2}|\cO_{K_D}^\times|$,  half the number of 
 units  of the  field $K_D$. This motivates the definition of a Heegner point in $E(K_D)$ as
 \beq
 P_{D,r} ~:=~ \frac{1}{u_D} \rmtr~ \Phi(z_{D,r})  ~\in ~E(K_D).
 \eeq

For any given discriminant $D$ there may or may not exist nontrivial points
that are defined over the field $K_D$ and the question arises
whether non-trivial D$-$crystals can be obtained for a given
elliptic curve. If so then the complexity of these arithmetic
D$-$crystals can be measured in terms of the N\'eron-Tate height
$\hhat(P_D)$ which defines a real valued function on $E(K_D)$.
This existence problem will be addressed next.

\subsection{The existence of $K_D-$type D$-$crystals}

The conjecture of BSD$_K$II shows that in order to obtain
nontrivial D$-$crystals whose base points take values in $K_D$ it is necessary to
have curves $E/K_D$ such that the derivative of their
$L-$functions does not vanish at the critical point.  The $L-$function of $E/K_D$
can be computed in terms of curves over the rational field by considering the twisted curve 
 $E^D$ of the the curve $E$
  \beq
  L(E/K_D,s) ~=~ L(E,s) ~L(E^D,s).
  \lleq{L-of-EKD}
  If the curve $E$ is described by a cubic polynomial $p(x)$ as 
    \beq
  E:~y^2 ~=~ x^3 + ax^2 + bx + c,
  \eeq
  its twist is defined as $E^D: ~Dy^2 =p(x)$, and can be brought into the standard cubic form by
    \beq
  E^D:~~y^2 = x^3 + aDx^2 + bD^2x + cD^3.
  \eeq
 
If the Hasse-Weil $L-$function of $E$ is expanded as 
  \beq
  L(E,s) = \sum_n a_n n^{-s}
  \eeq
  and $\chi_{D}(p)$ is again the character associated to 
   $K_D$ via the Legendre symbol,  the $L-$function of the twisted curve $E^D$ is given by twisting 
the coefficients $a_n$ by this character
   \beq
   L(E^D,s) = \sum_n a_n \chi_D(n) n^{-s} =: L(E,s) \otimes \chi_D.
  \eeq
  The conductor of the twisted curve $E^D$ for discriminants $D$ that are 
  coprime to the conductor $N(E)$ of are given by
 \beq
   N(E^D) = N(E)D^2,
  \eeq
 and the sign of the functional equation for the twisted curve is finally given
 in terms of the sign $\e=\e(E)$ of the original curve and the character $\chi_D$ by
 \beq
 \e(E^D) = \e(E) \chi_D(-N(E)).
 \eeq
 These relations make it possible to understand  the
 conjecture of Birch and Swinnerton-Dyer for $E/K_D$ in terms of
 the behavior of the curve $E$ and its twists $E^D$ defined over
 the rational number field.

 It follows from (\ref{L-of-EKD}) that the non-vanishing of the $L-$series at the critical point 
 can be guaranteed either by having (twisted)
curves with $L'(E,1)\neq 0\neq L(E^D,1)$ or via $L(E,1)\neq 0 \neq
L'(E^D,1)$. Information about the existence of such twisted curves
can be obtained by considering the behavior of twists of modular
forms of weight 2. This follows from the elliptic modularity
theorem of Wiles and Taylor, as well as Breuil-Conrad-Diamond-Taylor,
which says that for each elliptic curve $L(E,s)=L(f,s)$ where
$f\in S_2(\G_0(N))$. Furthermore, the $L-$function of $E^D$ is
given by $L(E^D,s) = L(f\otimes \chi,s)$.
 Results by Bump-Friedberg-Hoffstein \cite{bfh89, bfh90},
 Murty-Murty \cite{mm91} and Waldspurger \cite{w85} then imply the
 existence of infinitely many algebraic D$-$crystals. The details
 depend on the sign of the functional equation as follows.
Let $f\in S_{2w}(\G_0(N))$ be a cusp new-form with trivial
character and the completed $L-$function normalized as
 in eq. (\ref{completed-L-function})  which satisfies the functional equation
  (\ref{functional-equation}), 
 where $\e=\pm 1$. Heegner points exist for both signs of the
 functional equation.

 First, it was shown in \cite{bfh90,mm91} that if the sign $\e$ of the
 functional equation is positive then there exist infinitely many
 imaginary fields $\mathQ(\sqrt{-D})$ of discriminant $-D$
 prime to $N$ such that the Heegner constraint is satisfied, i.e.
 all prime divisors of $N$ split in $K_D$, and the twisted $L-$function 
$L(f\otimes \chi_D,s)$
 has a first order zero at $s=w$. If for $w=1$ the modular form
 $f\in S_2(\G_0(N))$ is such that
 $L(E,1) = L(f,1)\neq 0$ the existence of these twists therefore implies the
 nonvanishing of the height of the Heegner point associated to $K_D$.

 If the sign $\e$ is negative then $\rmord_{s=1}L(E,s) \geq 1$ and a result of Waldspurger
 \cite{w85} shows that there exist infinitely many quadratic
 characters $\chi_D$ such that the $L-$function of the twisted curve $E^D$ does not
 vanish, $L(E^D,1) \neq 0$. It follows from the work
 of Kolyvagin that the twisted curves $E^D$ have rank zero. Combining this
  result with the theorem  of Gross and Zagier again guarantees
  the nonvanishing of the height of the Heegner points associated to $K_D$.
 It has been conjectured by Goldfeld \cite{g79} that for
 newforms of weight two the twisted $L-$functions $L(f\otimes \chi_D,1)$ do
 not vanish for $\frac{1}{2}$ of the square-free integers $D$.

The results just described therefore show that for general elliptic curves there are an infinite 
number of discriminants $-D$ such that the Heegner points of $K_D$ define base points 
of algebraic D$-$crystals 
 \beq
  D_H(E) ~=~ \bigcup_{K_D} \mathZ P_D.
 \eeq
Such D$-$crystals can then be used as probes the compactification geometry as well as the 
 worldsheet theory, the latter mediated by the $L-$function.

\subsection{General Heegner points}

Less well understood, but also of interest for the physics of
D0$-$branes, are constructions of algebraic points on elliptic
curves that go beyond Birch's systematization \cite{b75, bs84} of
Heegner points. The aim of these generalization is to relax the
Heegner constraint on the imaginary quadratic extensions $K_D$
over which the elliptic curve is defined since it is of interest
to try to construct rational points over such $K_D$ for which the
prime divisors of the level $N$ does not necessarily split in $K$.
In the present paper this will not be described in a systematic
way, but will be exemplified for a particular curve which allows
not only to relate Heegner type D$-$crystals to the geometry of the
compactification manifold, but also to the structure on the string
worldsheet via the modular form associated to the motivic $L-$function.

\vskip .2truein

\section{Heights of $K_D-$type D$-$crystals and $L-$functions}

 The rank part of the conjecture of Birch and Swinnerton-Dyer
 $\rmBSD_\mathQ$I implies that for curves with positive rank
 the $L-$function at the central critical point vanishes, $L(E,1)=0$.
 It is therefore natural to ask what the role is of the derivative
 at this point. 

\subsection{The formula of Birch,  Gross-Zagier and Zhang}

In the 1960s Birch conjectured, based on
 numerical experiments, that the height of
Heegner points associated to imaginary quadratic fields $K_D =
\mathQ(\sqrt{-D})$ of discriminant $-D$ should be related to the
special value of the derivative of the Hasse-Weil $L-$function
 of $E/K_D$ evaluated at the central
 critical point \cite{bb04}
 \beq
 \hhat(P_D) = c_{E,D} L'(E/K_D,1),
 \lleq{birch-conjecture}
 where $K_D$ satisfies the Heegner condition and $c_{E,D}$ is a constant that
 depends on the elliptic curve $E$ and the field $K_D$, but was not completely specified 
by the experimental data.

 This conjecture was made precise by Gross and Zagier who proved the 
  formula conjectured 
by Birch (\ref{birch-conjecture}), in the process providing  an explicit form of the coefficient $c_{E,D}$.
 More precisely, for imaginary quadratic fields $K_D$ with odd $D$ that satisfy 
the Heegner condition it was shown in \cite{gz86} that 
   \beq
   L'(E/K_D,1) = \frac{1}{c^2\cdot u_D^2} \frac{||\om||^2}{\sqrt{D}}
    ~\hhat(P_D),
  \eeq
 where $\hhat(P_D)$ is the N\'eron-Tate height of the Heegner
 point $P_D \in E(K_D)$. The Manin constant $c$ relates the
 N\'eron differential $\om$ on $E$ to the differential
 $\om_f = 2\pi i f(z)dz$ on the modular curve $X_0(N)$ via the modular representation
  $\Phi_N^*(\om) = c\om_f$. It is expected that $c=1$, which
  has been shown for square-free $N$ in ref. \cite{e91}. 
 For the N\'eron differential $\om$ the norm $||\om||^2$ is defined as
 \beq
 ||\om||^2 = \int_{E(\mathC)} \om \wedge \overline{i\om}.
 \eeq
 Finally,
  $u_D$ denotes one half of the number  of units in $K_D=\mathQ(\sqrt{-D})$.
  One finds that $u_D=1$ except for the Gauss field $\mathQ(\sqrt{-1})$, for which 
 $u_1=2$, and the Eisenstein field $\mathQ(\sqrt{-3})$, for which $u_3=3$.
 The Gross-Zagier theorem was extended later for general discriminants by Zhang \cite{z04}.

The key consequence of the Gross-Zagier formula is that the point $P_D$ has
 infinite order if and only if the $L'(E/K_D,1)\neq 0$.

 Comparing the theorem of Gross and Zagier with the conjecture of
Birch and Swinnerton-Dyer leads to the expectation that the
subgroup of $E(K)$ generated by the Heegner point $P_D$ is related
to the order of the Shafarevich-Tate group as
\beq
  \left( c \prod_p c_p \right)^2 |\Sha_K(E)| ~=~[E(K_D):\mathZ P_D]^2.
\eeq

 The details of the Gross-Zagier relation depend on the type of
 the elliptic curve under consideration. In the present
paper the focus will be on the two most important situations which
together make up the vast majority of all the cases encountered
for elliptic curves. These types are characterized by the rank of
the Mordell-Weil group. It is not known whether the rank of this
group is bounded, but the existing databases all indicate that the
ranks 0 an 1 combined provide by far the majority of all known
curves. The following discussion therefore aims at this most
important class of curves.

\subsection{Curves with $\rmrk~E(\mathQ) =0$}

The elliptic curves of Brieskorn-Pham type considered above for
which the reconstruction from string worldsheet is completely
known \cite{rs05} have D$-$crystals with base points that are pure
torsion. For the case $\rmrk~E(\mathQ)=0$ the rank part of the
 conjecture of Birch and Swinnerton-Dyer implies that the $L-$function
 does not vanish at the central critical point, $L(E,1)\neq 0$.
  The conjecture of BSD$_{\mathQ}$II then implies that it
  essentially determines the period of the curve, as well as the order
  of the torsion group and the order of the
 Shafarevich-Tate group $\Sha$ via the specialized form
 (\ref{bsd2-rk0}).

 For the Brieskorn-Pham curves in particular, and rank 0 curves in
 general, the goal therefore is to find twists $E^D$ that lead to root
 numbers $\e(f\otimes \chi_D)=-1$, hence to vanishing
 $L-$functions. The existence of such $D$ is guaranteed by the result
 described in Section 4. The generator in these cases thus comes from $E^D$
 and the Gross-Zagier formula takes the form
 \beq
   \hhat(P_D) = \frac{u_D^2\sqrt{|D|}}{||\om_E||^2} L(E,1)L'(E^D,1)
 \eeq
 for $\rmrk~E(\mathQ) =0$.

This result can be read in two ways. For exactly solvable curves one might view the 
  modular form derived from the worldsheet conformal field theory as the primary 
physical object. Associated to this modular form is the $L-$function and the formula 
of Gross and Zagier can be interpreted as providing information about spacetime objects,
namely the height of D$-$crystals, in terms of the worldsheet theory, mediated by the 
$L-$function. Inverting this point of view one can consider D$-$crystals as probes of the 
geometry of spacetime to gain information about the worldsheet theory, again mediated by 
the $L-$function and its associated modular form.

\subsection{Curves with $\rmrk~E(\mathQ)=1$}

For curves $E$ of $\mathQ-$rank 1 the conjecture of Birch and
Swinnerton-Dyer predicts that the $L-$function vanishes at the
central critical point, $L(E,1)=0$. Hence the derivative of
$L-$function of $E$ over the imaginary quadratic field $K_D$
 takes the form $L'(E/K_D,1) = L'(E,1)L(E^D,1)$.
The Gross-Zagier formula therefore reduces to
 \beq
  \hhat(P_D) = \frac{u^2\sqrt{|D|}}{||\om||^2} L'(E,1)L(E^D,1).
 \lleq{rank1gz}
 It follows that the generator of $E(K_D)$ comes from the
 rational Mordell-Weil group $E(\mathQ)$.
 Recalling that $L(E^D,s) = L(f\otimes \chi_D,s)$ with
 $L(E,s)=L(f,s)$ shows that the height determines the elliptic
 curve $E$ itself. This will be made more precise below.
 It is therefore possible to compare the $K_D-$rational points
 with the generator  $P \in E(\mathQ)$, and to ask whether the
 integral multiples $c_D$ of
  \beq
   P_D = c_D P
  \eeq
 have a geometric interpretation. This turns out to be the
 case. The work of Gross-Kohnen-Zagier shows that the values $c_D$
 are the coefficients of a modular form $F = \sum_n c_n q^n$
 of weight 3/2 whose image
 under the Shimura map \cite{s73,k85}
 \beq
  \rmSh: ~S_{3/2}(N) ~\lra ~ S_2(\G_0(N))
 \eeq
 is the motivic form $f(E,q) \in S_2(\G_0(N))$ if $E$ has conductor
 $N$. By a result of Waldspurger the coefficients $c_D$ are
 determined by special values of the motivic $L-$function of the
 twisted curve $E^D$.

\section{Characterization of the compactification geometry via D$-$crystals}

An essential ingredient of the program described in \cite{rs08} (and
references therein) to construct spacetime in string theory via
arithmetic methods is that the information contained in
automorphic forms is discrete. Using worldsheet modular forms to
construct automorphic forms and automorphic motives works as a
tool to construct the geometry of spacetime works because the
discrete information contained in the associated motivic
$L-$function is sufficient to characterize the motive. The key
result thus is that in automorphic geometries probes of finite
sensitivity completely determine the geometry. Following the same
logic in the present context of D$-$branes motivates the question
whether $K-$arithmetic D$-$crystals can be viewed as probes that are
sufficiently sensitive to determine the geometry uniquely (up to
isogeny).

The results described in the context of the Brieskorn-Pham exactly
solvable elliptic curves suggest that in general $\mathQ$-rational
points on elliptic curves will not be sensitive enough to
determine the geometry. These curves are of rank 0, hence the
D$-$crystals are of torsion type, of which there are only 15
different types. This means that there is an enormous degeneracy
among all the elliptic curves of rank 0 as far as their
$\mathQ-$rational structure is concerned. The next logical step is
to consider Heegner type D$-$crystals because they involve the
simplest possible generalization of $\mathQ$.

It is known that the geometry of an elliptic curve is determined
by its Hasse-Weil $L-$function up to isogeny, a result that was
proven by Faltings \cite{f83}. The relation between the height of
D0$-$branes and special values of $L-$functions raises the question
whether it is possible to characterize the geometry also in terms
of the special class of Heegner type D$-$crystals associated to
imaginary quadratic fields $K_D=\mathQ(\sqrt{-D})$, where $D$ is
viewed as a variable parameter. This is in fact possible by
interpreting a result by Luo-Ramakrishnan \cite{lr97} as a statement about 
 D$-$crystals.

{\bf Theorem.}~{\it Let $E,E'$ be two elliptic curves over
$\mathQ$ of conductors $N,N'$ such that for a non-zero scalar $C$
one has}
 $$
 \hhat(c_{f,K}) = C \hhat(c_{f',K})
 $$
 {\it where $f,f'$ are the corresponding modular forms and
 $c_{f,K}, c_{f',K}$ are the Heegner divisor components in
 $J(K)\otimes \mathQ$ for imaginary quadratic fields $K$. Then
 $N=N'$ and $E$ and $E'$ are isogenous over $\mathQ$.}

The proof of this result is based on \cite{lr97a}. This theorem
 shows that the Heegner type structure is sensitive enough to
 probe differences between elliptic curves at the level of
 isogenies.
The theorem is precisely the result needed to see that the D$-$crystals of Heegner 
type are sufficiently sensitive to probe the compactification manifold in the needed detail
because the $L-$function is an isgogeny invariant.  Since the $L-$function is determined 
for the exactly solvable curves by the modular forms on worldsheet it follows that the 
manifold is determined by the conformal field theory up to isogeny, as shown in 
\cite{su02, rs05}.

\section{Examples}

In this penultimate section two examples are discussed that
illustrate the behavior for both elliptic curves $E$ that are of
rank 0 and 1 respectively. The curves for which the relation
between spacetime geometry and worldsheet physics has been
analyzed so far are of the former type, while for rank 1 no such
relation is known at present. It is therefore of interest to show
that the context of D$-$branes does provide another route from
string theoretic objects to the reconstruction of the
compactification geometries. The two examples are motivated by the
choice of an example with known string worldsheet construction
\cite{rs05} and an example for which no such interpretation is
known. Many other examples can be analyzed similarly by making use
of the databases of Cremona, Stein and others.

\subsection{The exactly solve curve $E_{64}$}

As an example of a Brieskorn-Pham type elliptic curve consider
 the curve embedded in the weighted projective plane $\mathP_{(1,1,2)}$.
 This is a curve of conductor $N=64$ and in affine coordinates takes the form
  \beq
  E_{64}:~~y^2 = x^3 + x.
  \eeq
 For this example the precise relation is known between the
 modular form $f(E_{64},q) \in S_2(\G_0(64))$ associated to its
 Hasse-Weil $L-$function $L(E_{64},s)$ and the modular form on the
 string worldsheet, as noted in Section 2. The latter is a Hecke
 indefinite modular form
 $\Theta^2_{1,1}(\tau) = \eta(\tau)\eta(2\tau)$ and the relation involves
 a twist via the quadratic character associated to its field of
 quantum dimensions of the underlying conformal field theory
 \cite{rs05}
 \beq
  f(E_{64},q) ~=~ \Theta^2_{1,1}(4\tau)^2 \otimes \chi_2.
 \lleq{e64-relation}
 The curve $E_{64}$ has $\mathQ$-rank 0, hence the base points of the $\mathQ$-rational
 D$-$crystals are pure torsion. In order to obtain non-torsion type crystals it is
 useful to consider the curve $E$ over extensions $K$ of the rational numbers.

Heegner type D$-$crystals obtained on $E/K_D$ via imaginary
quadratic fields $K_D$ of discriminant $D$ involves the family of
twisted curves $E_{64}^D: ~Dy^2 = x^3 +x$, which can alternatively
be written in the standard Weierstrass form as
\beq
 E_{64}^D: ~y^2 ~=~ x^3 + D^2 x.
\eeq
 For discriminants $D$ such that $(N,D)=1$ the conductor of the
 twisted curve is given by $N(E_{64}^D) ~=~ 64D^2$,
 and the question then becomes for which conductor the twist curve
 has rank 1. It should be noted that the standard Heegner
 construction of Birch does not detect the existing rational
 points in this case. Direct computation (in part with the help of
 Stein's database) leads to the results of Table 3, which contains
 enough information to establish the relations of Birch and
 Swinnerton-Dyer in this case.
 
\begin{small}
 \begin{center}
 \begin{tabular}{c| c c c c c}
 $D$                          & &3               &7             &11          &19  \tabroom \\
 \hline
 $N(E_{64}^D)$          & &576            &3136          &7744        &23104  \tabroom \\

 $h(K_D)$                  & &1                &1             &1           & 1   \tabroom \\

 $\Om_{E_{64}^D}$   & &2.140901... &1.401548...   &1.1180...   & 0.850707...   \tabroom \\

 $R_{E_{64}^D}$        & &1.777251... &4.268241...   &8.5062...   &14.457238...   \tabroom \\

 $L'(E_{64}^D,1)$      & &1.902460... &2.991074...   &4.7552...   & 6.149442...    \tabroom \\

 $|E_{64}(\mathQ)_{\rmtor}|$
                                &  &2               &2              &2           & 2   \tabroom \\

 $c_p$                     &   &1,2            &2             &2           & 2   \tabroom \\

 \hline

 \end{tabular}
 \end{center}
 \end{small}
 \centerline{{\bf Table 3.} ~{\it Rank 1 twists in the family of
 curves $E_{64}^D$.}}

 This then shows that the heights of
 the rational D0$-$branes on the elliptic curve $E_{64}$ are given
 in terms of the $L-$function $L(E_{64},s)$, which in turn leads to
 the string theoretic modular form on the worldsheet via eq.
 (\ref{e64-relation}).

\subsection{The rank 1 curve $E_{37}$}

This section exemplifies the strategy of constructing the
compactification geometry from the structure of CM D0$-$branes in a
simple case. Consider the elliptic curve defined in its affine
form by
 \beq
  E_{37}: ~~y^2 + y = x^3 -x,
 \eeq
 which has conductor $N=37$ and which is birationally equivalent to
 the Weierstrass form 
 \beq
  E_{37}^W: ~~y^2 = x^3-16x+16.
 \eeq
 This curve is not of Brieskorn-Pham type and goes beyond the framework
 discussed in \cite{rs05}.

 The rank of the Mordell-Weil group is 1, in agreement with the
 prediction of the BSD-I formula since the sign of the functional equation
 of the modular for associated to the Hasse-Weil L-series
  \beq
   S_2(\G_0(37)) \ni f_{37}(q) ~=~
    q -2q^2 -3q^3 -2q^5 - q^7 -5q^{11} - 2q^{13} +2q^{23} +6q^{29}
    + \cdots
 \eeq
 is $\e=-1$ and 
  \beq
 L(E_{37},s) ~=~ L(f_{37},s).
 \eeq

 The Mordell-Weil group has no torsion,
 hence the group of rational points is given by
 \beq
  E(\mathQ) \cong \mathZ,
 \eeq
 with generator $P_\rmgen=(0,0)$. This example has been discussed in some detail by Zagier
 \cite{z84} and much of the necessary data for this curve can also
 be obtained from the tables constructed by Cremona \cite{c97}.
 The remaining BSD ingredients of this curve are given by
\bea
  R_{E_{37}}                    &=& 0.051111... \nn \\
  \Om_{E_{37}}^\mathR &=& 2.99345... \nn \\
  L'(E_{37},1)                 &=& 0.305999... \nn \\
    c_{37} &=& 1,
 \eea
 which leads to the analytic rank 1 of the Shafarevich-Tate group.
 The complex periods of  this curve is given by $\Om_{E_{37}}^2 = 2.451389...i$,
   leading to the norm of the N\'eron differential
  $||\om||^2 = 7.33813...$

The twists  $E_{37}^D$ of $E_{37}$ can be described in a
Weierstrass form as the family of curves
 \beq
  E_{37}^D: ~~y^2 = x^3 - 16D^2 x + 16D^3.
 \eeq

The Heegner condition $\chi_D(N)=1$ can easily be computed and
for the solutions it is possible to use the Gross-Zagier relation
for the rank 1 case (\ref{rank1gz}) to relate the $L-$function
values to the height of the resulting
 Heegner points. Table 4 illustrates this situation for a few
 Heegner points.
 \begin{center}
 \begin{tabular}{c | c c c c c}

 $D$                    & &7           &11         &47          &67     \\
 \hline

 $N(E_{37}^D)$    & &1813    &4477      &81733     &166093    \tabroom \\

 $L(E_{37}^D,1)$ & &1.853... &1.475...  &0.715...   &21.562...  \tabroom \\

 $\hhat(P_D)$     & &0.204... &0.204...  &0.204...   &7.360...
 \tabroom \\

 \hline
 \end{tabular}
 \end{center}
\vskip .05truein

 \centerline{{\bf Table 4.}~{\it Results for Heegner points $P_D$ on $E_{37}/K_D$.}}

\section{Outlook}

This paper has shown that the notion of D$-$crystals as discrete
D$-$brane configuration based on number fields $K$ is useful as a
probe that to detect the physical structure of the
compactification manifold. These D$-$crystals are linked to the
motivic $L-$function of the manifold, which according to the
Langlands reciprocity conjecture is expected to  lead to automorphic
forms. In several cases where this automorphic form is known it
was shown that the underlying motive is in fact modular that the
corresponding modular form encodes the physics of worldsheet via
modular forms that arise from the underlying conformal field
theory.

The results of this paper raise the question whether the strategy
described here can be generalized to higher dimensions. While much less
is known in this case, and even less has been proven, it is possible to indicate 
the framework into which higher dimensional D$-$crystals could be placed. On the
motivic side there are concrete examples of K3 surfaces
and Calabi-Yau threefolds of weighted hypersurface type 
for which motivic modular forms have been derived from the
motivic $L-$functions \cite{rs06, rs08}. For diagonal hypersurfaces 
these motivic $L-$functions and their automorphic forms are related to 
CFT theoretic forms on the worldsheet, hence the discussion of the 
exactly solvable elliptic curves in this paper should generalize. 
 Furthermore, there exist a
framework that generalizes the Mordell-Weil groups to higher
dimensions, and which allows to formulate conjectures that
generalize the conjectures of Birch$-$Swinnerton-Dyer and the
results of Gross and Zagier.

The higher dimensional conjectures generalizing the BSD framework
are formulated in terms of certain types of algebraic cycles,
which can be viewed as base schemes with D$-$branes wrapped around
them. The analog of the rank conjecture in this framework is the
conjecture of Beilinson and Bloch, which relates again the rank of
a particular group, the Chow group $\rmCH^r(X)_\rmhom$ of
null-homologous algebraic cycles of codimension $r$, to the
vanishing order of the $L-$function assigned to the intermediate
cohomology of the variety
 \beq
  \rmord_{s=r}L(H^{2r-1}(X),s) ~=~ \rmrk~\rmCH^r(X)_\rmhom.
 \eeq
 For $r=1$ the Chow group $\rmCH^1(X)_\rmhom$ of codimension 1 algebraic cycles describes
 points on the elliptic curves, and the Beilinson-Bloch conjecture
 reduces to the rank conjecture of Birch and
 Swinnerton-Dyer.

Embedded in the Chow group of cycles one can consider
generalizations of the Heegner points, in this case Heegner
cycles, or more generally complex multiplication
 cycles. In the case of CM cycles no proof exists for the
 analog of the Gross-Zagier theorem in this case, but the conjectural
 framework can be used for guidance. It would be of great interest
 to generalize the framework of the present paper to the context
 of higher dimensional D$-$crystals. Furthermore, it would be of
 importance to generalize this framework from a
 cohomological to a motivic setting, in which the $L-$functions
 considered are that of motives, i.e. the lhs is concerned with
 the vanishing order of $L(M,s)$, with $M$ the motive.

\vskip .3truein

{\bf Acknowledgement.}

It is a pleasure to thank Ulf K\"uhn and Monika Lynker  for
conversations. This
work has been conducted over a number of years, during which I
have benefitted from the support and hospitality of several
institutions. First, I thank the Mathematics Institute Oberwolfach
for support during an extended stay. Some of the results of this
paper were presented at the Banff International Research Station and I thank
Chuck Doran, Noriko Yui and Don Zagier for the opportunity to
present this work. Visits to the Werner Heisenberg Max Planck Institute in 
 Munich and to CERN greatly facilitated the conclusion of this project and 
 I thank the Max Planck Gesellschaft 
and CERN for their support.  Thanks are due in particular to Dieter L\"ust in Munich and Wolfgang Lerche 
at CERN  for making these visits possible, and I thank the string theory groups at both 
  institutions for their friendly hospitality. This work was
supported in part by an IUSB Faculty Research Grant and a grant by the National Science 
Foundation under grant No. 0969875.

 \vskip .2truein



\end{document}